\documentclass[runningheads]{llncs}
\usepackage{graphicx}
\usepackage{subcaption}
\usepackage{amsmath}%
\usepackage{amssymb}

\begin{document}
\title{Your Favorite Gameplay Speaks Volumes about You: Predicting User Behavior and Hexad Type\thanks{The authors' pre-print version of the paper accepted at the conference of HCI International 2023}}
%
%
\author{Reza Hadi Mogavi\inst{1}\and
Chao Deng\inst{2}\and Jennifer Hoffman\inst{2}\and Ehsan-Ul Haq\inst{1}\and Sujit Gujar\inst{3}\and Antonio Bucchiarone\inst{4}\and Pan Hui\inst{1, 5}}
\authorrunning{Reza H. Mogavi et al.}
%
\institute{Hong Kong University of Science and Technology, Hong Kong SAR\\\email{\{rhadimogavi, euhaq\}@connect.ust.hk, panhui@ust.hk}
\and
Accessible Meta Research and Development, Michigan, United States\\\email{\{cdeng, jhoffman\}@accessiblemeta.org}
\and
International Institute of Information Technology, Hyderabad, India\\\email{sujit.gujar@iiit.ac.in}
\and
Fondazione Bruno Kessler, Trento, Italy\\\email{bucchiarone@fbk.eu}
\and
University of Helsinki, Helsinki, Finland\\\email{pan.hui@helsinki.fi}
}
\titlerunning{Your Favorite Gameplay Speaks Volumes about You}

\maketitle              
\begin{abstract}
In recent years, the gamification research community has widely and frequently questioned the effectiveness of one-size-fits-all gamification schemes. In consequence, personalization seems to be an important part of any successful gamification design. Personalization can be improved by understanding user behavior and Hexad player/user type. This paper comes with an original research idea: It investigates whether users' game-related data (collected via various gamer-archetype surveys) can be used to predict their \textit{behavioral characteristics} and \textit{Hexad user types} in non-game (but gamified) contexts. The affinity that exists between the concepts of \textit{gamification} and \textit{gaming} provided us with the impetus for running this exploratory research. 

We conducted an initial survey study with 67 Stack Exchange users (as a case study). We discovered that users' gameplay information could reveal valuable and helpful information about their behavioral characteristics and Hexad user types in a non-gaming (but gamified) environment. 

The results of testing three gamer archetypes (i.e., \textit{Bartle}, \textit{Big Five}, and \textit{BrainHex}) show that they can all help predict users' most dominant Stack Exchange behavioral characteristics and Hexad user type better than a random labeler's baseline. That said, of all the gamer archetypes analyzed in this paper, \textit{BrainHex} performs the best. In the end, we introduce a research agenda for future work.

\end{abstract}
\section{Introduction}
Over the years, games have found an integral place in the daily lives of many people around the globe \cite{Ryan2006,Granic2014,10.1145/3313831.3376755,Barr2021}. They can be found everywhere, from people's homes to their phones and, more recently, even in the metaverse \cite{Haberlin2022,Parry2014,paulmetaverse}. They offer endless opportunities for personal development \cite{Connolly2012,Galleguillos2019}, creativity \cite{10.1145/3311350.3347182}, and fun \cite{Cole2007}. According to the Entertainment Software Association's most recent survey, conducted in February 2022, more than 200 million Americans (nearly two thirds) regularly play at least one type of video game to bring joy and happiness into their lives \cite{ESA_Survey2022}. 

However, not all gamers enjoy the same types of games (or gameplays) \cite{10.1145/2470654.2481341,10.1145/3025453.3025577,10.1007/978-3-030-29384-0_23}. In fact, player preferences can vary widely based on various individual and contextual factors such as \textit{personality}, \textit{age}, \textit{gender}, \textit{occupation}, \textit{daily life schedule}, \textit{cultural background}, \textit{expenses}, and \textit{accessibility} \cite{10.1007/978-3-030-29384-0_23,Chesham2017,Boggio2020,thayer2004localization,Cairns2019}. Fortunately, however, there are now several applicable gamer models (or archetypes) in the literature that have proven helpful in describing \textit{parts of} these various complexities using simple (and reasonably easy-to-conceptualize) player attributes. 

Three of the \textbf{most frequently cited archetypes} of gamers in the human-computer interaction (HCI) and gameplay research communities (e.g., \textit{CHI Play} \cite{10.1145/3568732}) are \textit{Bartle} \cite{bartle1996hearts}, \textit{Big Five} \cite{deHesselle2021}, and \textit{BrainHex} \cite{Nacke2014}. 

Bartle's taxonomy of player types, for example, states that players can be divided into four groups based on how they like to play games: \textit{achievers}, \textit{socializers}, \textit{explorers}, and \textit{killers}. In brief, \textit{achievers} like to rack up achievements and badges; \textit{socializers} favor interacting with other players; \textit{explorers} favor gaining knowledge about the game world and its mechanics; and finally, \textit{killers} favor lurking and competing against other players.\footnote{More information can be found in the Section of Literature Review.} 

While such gamer archetypes may strike some scholars (and readers) as oversimplifications, they do a good job of providing a \textit{holistic overview} of some of the most common gamer interests that developers and game designers should be aware of \cite{10.1145/3311350.3347167,hamari2014player}. Consequently, it is witnessed that many giant game companies (e.g., \textit{Nintendo}, \textit{Ubisoft}, and \textit{Blizzard}), research institutes, and interest groups worldwide spend hundreds to thousands of dollars annually to collect such data from gamers through a variety of survey methods.

$\triangleright$ \textbf{Our research here presents an exploratory study that seeks to understand what else can be inferred from such survey data on people's game (or gameplay) interests. More specifically, we want to investigate whether gamer archetypes (such as Bartle) can tell us anything helpful about user behavior in \textit{non-game contexts} and \textit{Hexad user types}.}

\subsection{Background}
\textit{Gamification} has recently become an independent field of study in the HCI communities, and it historically has strong ties to the \textit{gaming industry}. The term gamification is \textit{often} defined as the use of game design elements (such as badges, points, and leaderboards) in non-game contexts (such as education, business, health, and crowdsourcing) \cite{10.1145/2181037.2181040,Seaborn2015,Hamari2014Work}. According to the extant literature, gamification has the ability to change user behavior \cite{10.1145/3178876.3186147,10.1145/2488388.2488398} and can have both positive (e.g., \cite{10.1145/3491140.3528274,Legaki2020}) and negative effects (e.g., \cite{10.1145/3491140.3528274,10.1145/3555553,10.1145/3555124}) on user activities. This relevant prior knowledge motivates us to explore the possibility of using players' game-related data to predict their most dominant user behavior in gamified (but non-game) environments.

\textbf{Beyond User Behavior.} Aside from user behavior, another vital topic worth discussing in any gamified environment is determining users' gamification preferences. Here, we are curious to find out whether game-related data can also be used to predict users' gamification preferences. To investigate users' gamification preferences more systematically (and practically), we employ \textit{Hexad}, which is a well-established empirical framework often used to identify and understand users' gamification preferences (or \textit{Hexad type}) \cite{10.1145/3173574.3174009,Lopez2019,10.1145/3311350.3347167}. 

\textbf{Hexad Framework.} According to the Hexad framework, different Hexad types include (1) \textit{Philanthropists} (provoked by: purpose), (2) \textit{Socializers} (provoked by: relatedness), (3) \textit{Free Spirits} (provoked by: autonomy), (4) \textit{Achievers} (provoked by: competence), (5) \textit{Players} (provoked by: extrinsic rewards), and (6) \textit{Disruptors} (provoked by: their urge to change).

In recent years, an increasing number of HCI researchers have explored many inventive methods for predicting Hexad types other than the typical surveying approach that is developed by Tondello et al. \cite{10.1145/2967934.2968082}. Each method has its own unique benefits and drawbacks and, when necessary, can be used in a variety of situations and contexts. For example, a research conducted by Altmeyer et al. showed that user-generated smartphone data can help predict Hexad types and potentially tailor gamification without requiring direct user interaction \cite{10.1145/3341215.3356266}. Nevertheless, as is evident, some people may have privacy and security concerns about sharing their smartphone data with gamification designers and developers. In another study, Kimpen et al. conducted an expert consensus study with a group of 11 experts and argued that banking data could as well be helpful in predicting Hexad types \cite{10.1145/3450337.3483486}. \textbf{With this background in mind, our research presents a new way to predict gamers' Hexad types based on the information about their game and gameplay preferences when such information is available. } 
\begin{figure}[t!]
    \centering
    \begin{minipage}[b]{0.70\textwidth}
        \centering
        \includegraphics[width=\textwidth]{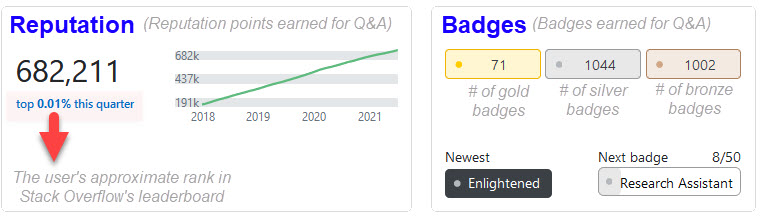}
        \subcaption{Reputation Points and Badges}~\label{fig:HMMTCH1}
    \end{minipage}\hfill
    \begin{minipage}[b]{0.29\textwidth}
        \centering
        \includegraphics[width=\textwidth]{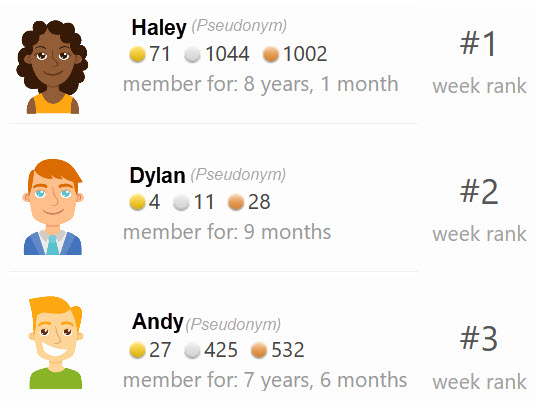}
        \subcaption{Leaderboard}~\label{fig:HMMTCU2}
    \end{minipage}
    \caption{Anonymized snapshots of Stack Exchange's gamification mechanisms: reputation points, badges, and the leaderboard}~\label{fig:SnapshotsSO}
\end{figure}
\subsection{Research Context}
The non-game context chosen for investigation in this article is \textbf{Stack Exchange}, which is a popular example of \textit{Community Question Answering Websites (CQAs)}. CQAs such as Stack Exchange, Quora, and Zhihu are among the most important Computer Supported Cooperative Work (CSCW) applications that have been in the interest scope of the HCI and gamification research communities for a long time \cite{10.1145/3555124,10.1145/3555553,10.1145/2934687}. These platforms help millions of people worldwide ask and answer questions online and serve as helpful knowledge repositories on different topics \cite{10.1145/2934687}. As shown in Figure 1, CQAs typically use various gamification elements such as badges, points, and leaderboards to incentivize user participation and prevent churn (i.e., user dropouts) \cite{rezaMogavi2019}.

\subsection{Key Contributions}
In summary, our \textbf{main contributions} to the existing body of knowledge in this area are as follows:
\begin{itemize}
    \item[\textbullet] According to our findings, there is a notable correlation between people's preferences for specific types of gameplay and their behavioral characteristics in non-game (but gamified) contexts, as well as their preferences for specific types of gamification.
    \item[\textbullet] According to our results, all player archetypes can (to some extent) help predict user behavior in Stack Exchange (a non-game gamified context) and their Hexad type better than the baseline of a random labeler. However, of all the gamer archetypes studied in this paper, \textit{BrainHex} is the one that has the best performance.  
    \item[\textbullet] Our research can help eliminate (or mitigate) the need for parallel surveys of people's preferences about games and gamification. This study shows that game-related data can be used to make reasonable inferences about people's dominant behavioral attributes and preferences for gamification. Thus, our research findings could help game/gamification researchers and survey respondents save time, money, and energy.
    \item[\textbullet] Last but not least, our findings encourage HCI researchers and practitioners to think outside the box and apply their existing knowledge resources/tools (such as player archetypes) to solve broader and different types of problems (e.g., predicting Hexad user type). In the ever-expanding HCI community of today, finding creative and novel solutions to domain-related issues appears to be quite essential.
\end{itemize}
\section{Literature Review}
This section reviews the literature in the areas of gamer archetypes, gamification, and Hexad user/player types.
\subsection{Gamer Archetypes}
Personalization and customization of the gaming experience are essential to ensure that gamers have the most fun and enjoy their time. Three of the most commonly cited gamer archetypes used for this purpose in the HCI and gameplay research communities are \textit{Bartle} \cite{bartle1996hearts}, \textit{Big Five} \cite{deHesselle2021}, and \textit{BrainHex} \cite{Nacke2014}. A great deal can be gleaned about gamers from each of these archetypes.\footnote{In this text, the terms \textit{typology}, \textit{archetype}, \textit{model}, and \textit{behavioral traits} are sometimes used interchangeably.}

$\Box$ \textbf{Bartle's Taxonomy of Player Types.} According to the gamer taxonomy developed by Bartle in 1996, gamers who play in multiplayer virtual worlds can be divided into four types: \textit{achievers}, \textit{socializers}, \textit{explorers}, and \textit{killers}. As shown in Figure \ref{fig:bartle}, these categories are determined by classifying gamer preferences along two dimensions, i.e., players-world spectrum (\textit{horizontal axis}) and acting-interacting spectrum (\textit{vertical axis}). Each gamer type is defined as follows \cite{reiners2015gami}:

\begin{figure}[t!]
        \centering
        \includegraphics[width=0.55\textwidth]{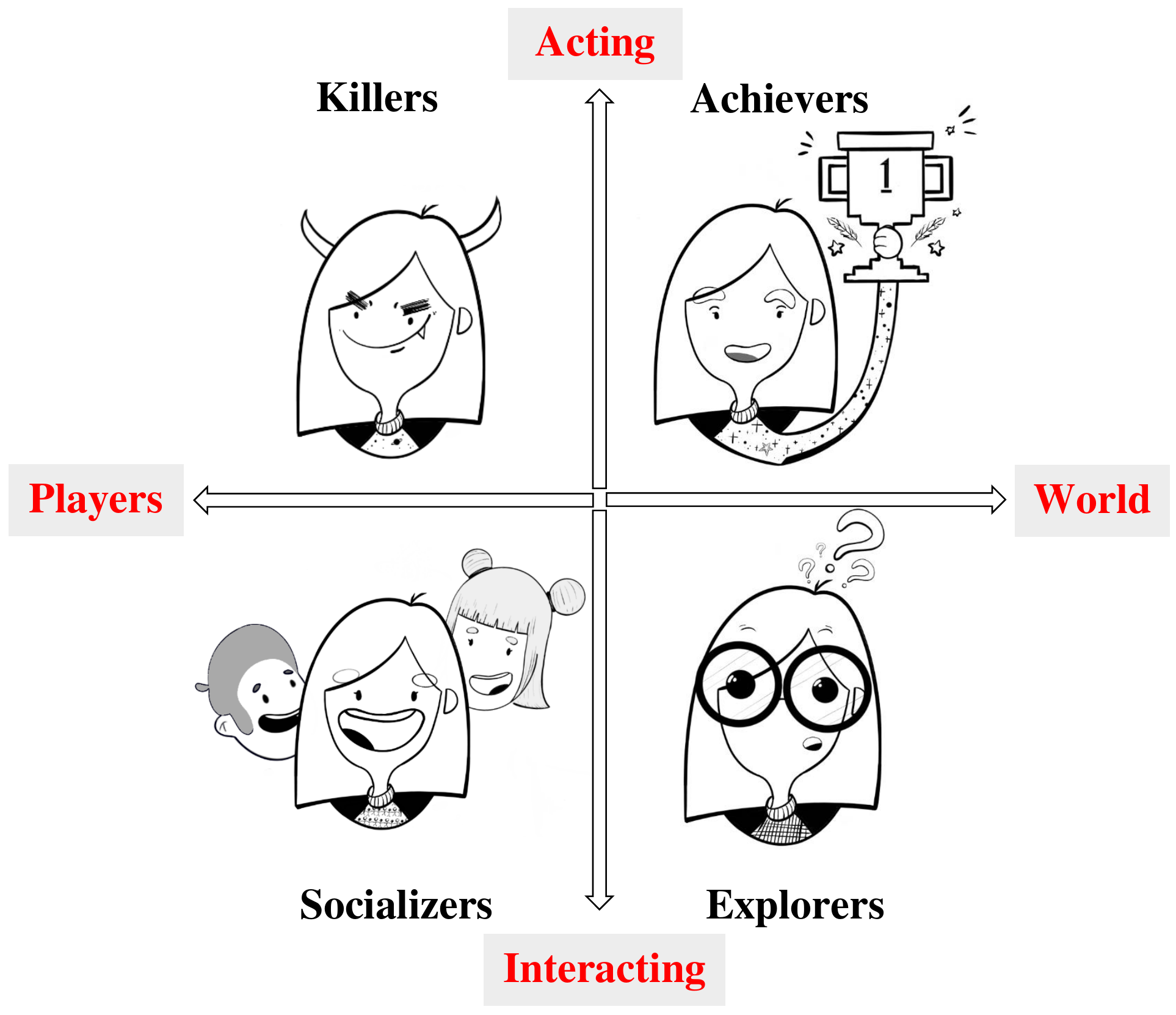}    \caption{Bartle's taxonomy of gamer types: Achievers, Socializers, Explorers, and Killers}~\label{fig:bartle}
\end{figure}

\textbf{Achievers} are defined as those who like to \textit{act in the world}. They are gamers who want to feel that they have achieved something significant (often scarce) in the game; they place great value on the element of \textit{competence} described in the \textit{Self-Determination Theory (SDT)}. According to this theory, people need a sense of \textit{competence} (being skilled and effective), \textit{relatedness} (being connected to others), and \textit{autonomy} (being master of their behavior) in order to be satisfied enough to perform certain actions \cite{10.1145/3313831.3376723,10.1145/3555124}.

\textbf{Explorers}, on the other hand, like to \textit{interact with the world}. They are gamers who want to engage with the full scope of the system and explore its boundaries and hidden secrets; they naturally value the element of \textit{autonomy} in self-determination theory the most.

\textbf{Socializers} enjoy \textit{interacting with other players}. They are gamers who want to use the system to meet new people and make new friends; socializers, according to the self-determination theory, are interested in the element of \textit{relatedness}. Finally, \textbf{killers} are gamers who take pleasure in doing things to other people or \textit{acting on other players}. They enjoy competing with others, and the \textit{competence} element of the self-determination theory appeals to them the most.

Currently, the most common empirical method for determining an individual's Bartle type is Dr. Matthew Barr's survey method, known as Bartle test.\footnote{http://matthewbarr.co.uk/bartle/}

$\Box$ \textbf{Big Five Personality Traits.}\footnote{Also called \textit{Five-Factor Model of Personality}} In psychology, personality traits go beyond the game/gamification world \cite{10.1145/3311350.3347167} and are understood as patterns of \textit{thought}, \textit{feeling}, and \textit{behavior} that are relatively \textbf{persistent} throughout a person's life \cite{BritWiki}. Developed and disseminated primarily in the 1980s and 1990s, the model of Big Five personality traits include
\textit{conscientiousness}, \textit{agreeableness}, \textit{neuroticism}, \textit{openness}, and \textit{extraversion} \cite{10.1145/3311350.3347167}. Below, we summarize the main characteristics of each personality trait:

\begin{itemize}
    \item \textbf{Conscientiousness} shows the proclivity for self-control, discipline, tenacity, organization, and responsibility toward others.
    \item \textbf{Agreeableness} indicates the emotional feelings such as empathy and concern for the needs of others, altruism, trust, and cooperation.
    \item  \textbf{Neuroticism} describes a person's proclivity to be under psychological pressure (instability) and to experience negative emotions such as depression, anxiety, and anger.
    \item \textbf{Openness}\footnote{Also called \textit{Openness to Experience}} describes a person's imaginative, creative, curious, risk-taking, and adventurous tendencies.
    \item \textbf{Extraversion (or extroversion)} describes a person's tendency to be gregarious, friendly, warm, and enthusiastic about activities that involve interaction with others, such as making new friends or participating in group projects.
\end{itemize}
TIPI (Ten Item Personality Measure) is currently the most widely used method for measuring a person's Big Five personality traits \cite{Gosling2003}.

$\Box$ \textbf{BrainHex Player Typology.} The BrainHex typology was developed in the early 2010s. It was inspired by several sources and references, including \textit{earlier typology approaches}, \textit{discussions of play patterns}, the \textit{literature on gameplay emotions}, and most importantly, the relevant \textit{neurobiological studies} \cite{bateman2011player,Nacke2014}. This typology divides gamers' playing motivations into seven categories \cite{Nacke2014}: \textit{Seeker} (motivated by exploration), \textit{Survivor} (driven by fear), \textit{Daredevil} (motivated by excitement), \textit{Mastermind} (motivated by strategy), \textit{Conqueror} (motivated by challenge), \textit{Socializer} (motivated by social relationships), and \textit{Achiever} (motivated by goal achievement). In the following, we summarize and explain each of these categories based on the explanations provided by Lennart Nacke et al. \cite{Nacke2014}:
\begin{itemize}
    \item \textbf{The Seeker} loves to explore the environment and have adventures. Endormorphin is released when a seeker's brain is exposed to rich patterns of interpretable information (often sensory), activating the pleasure center in their brain. 
    \item \textbf{The Survivor} takes pleasure when exposed to a high level of suspense induced by fear or anticipation of frightening situations, at least in the context of fictional activities (e.g., horror movies and video games). Among survivors, the arousal neurotransmitter epinephrine amplifies the impact of dopamine triggered by the receipt of rewards. However, it still needs to be determined whether the pleasure of fear should be evaluated by the intensity of the fright itself or by the relief felt afterward.
     \item \textbf{The Daredevil} revels in the thrill and peril of playing on the edge and taking risks (e.g., playing at extreme heights or participating in a car race at extreme speeds). Similar to Survivors, epinephrine is a reward enhancer in daredevils. 
    \item \textbf{The Mastermind} takes pleasure in solving complex puzzles, developing clever strategies, and making the most efficient decisions. The pleasure and decision centers of a mastermind's brain are closely connected, and good decisions are inherently rewarded. The mastermind archetype is very similar to Bartle's explorer type. 
    \item \textbf{The Conqueror} describes a challenge-seeking gamer who is not content with easy victories. Difficult situations stimulate the release of epinephrine (adrenalin), associated with arousal and excitement, and norepinephrine, associated with anger. Regardless of gender, testosterone is also thought to play a role in this archetype. According to the literature, the conqueror archetype is weakly related to Bartle's killer type.
    \item \textbf{The Socializer} enjoys interacting with others, hanging out with trusted people, talking with them, and assisting them. This archetype is linked to the social center of the brain, which is the primary neural source of oxytocin, a neurotransmitter associated with trust. The name of this archetype alludes to Bartle's socializer archetype.
    \item \textbf{The Achiever} describes a goal-oriented gamer motivated by long-term achievements. In this archetype, the release of dopamine and stimulation of the brain's pleasure center contribute to a sense of personal fulfillment. The name of this archetype alludes to Bartle's achiever archetype.
\end{itemize}
Currently, most HCI and gameplay researchers and practitioners use the official \textit{International Hobo Questionnaire} \cite{IHBO} to determine the BrainHex types of their research participants.
\subsection{Gamification and Hexad}
In recent years, gamification has emerged as an essential research topic in the fields of HCI, CSCW, and Learning at Scale. Simply put, gamification is the application of game design elements such as \textit{badges}, \textit{points}, and \textit{leaderboards} in non-game contexts \cite{10.1145/2181037.2181040}. However, the mere addition of badges, points, and leaderboards to a system/service does not guarantee user engagement or lead users toward a desired behavior \cite{10.1145/3555553,10.1145/3555124,10.1145/3491140.3528274}. According to Nicholson, a good gamification design should support its target users in making \textit{meaningful} connections between their personal goals and the use of gamification \cite{Nicholson2014}. This is important because it can help improve users' behavioral outcomes and experiences. The literature suggests that a tailored gamification design can help promote the formation of such perceptual connections \cite{Nicholson2014,10.1145/3311350.3347167,10.1145/3555124}.

$\Box$ \textbf{Hexad} \cite{10.1145/2967934.2968082} is a framework widely used by the gamification research community to adapt gamification in various systems and applications. It is based on the \textit{self-determination theory (SDT)} \cite{10.1145/3313831.3376723}. 
Hexad framework helps gamification researchers and designers understand users' \textit{motivations} for interacting with different types of gamification elements in different environments. Below, we review the main characteristics of each Hexad type (according to \cite{10.1145/2967934.2968082}):
\begin{itemize}
    \item \textbf{Philanthropists} seek \textit{purpose} and \textit{meaning}. They are altruistic and willing to help or give without expecting anything in return.
    \item \textbf{Socializers} are motivated by \textit{relatedness} (as described in SDT). They want to interact with others and form social bonds.
    \item \textbf{Free Spirits} are motivated by \textit{autonomy} (as described in SDT). They enjoy creation and exploration without any controls or restrictions.
    \item \textbf{Achievers} seek \textit{competence} (as described in SDT). They want to prove their worth by taking on challenging tasks and moving up in a system.
    \item \textbf{Players} are after the \textit{extrinsic rewards}. They are willing to engage in any activity (regardless of the type) in order to receive a reward within a system.
    \item \textbf{Disruptors} are driven by their will to initiate a \textit{change} (either positive or negative) in the system. They enjoy challenging the system.
\end{itemize}
The survey approach developed by Tondello et al. \cite{10.1145/2967934.2968082} is the most common method for determining a user's Hexad type. However, as the Hexad framework becomes more popular, new approaches to predicting the Hexad types are being developed that can be useful in a variety of situations based on the availability of different types of data \cite{10.1145/3410404.3414232,10.1145/3341215.3356266,10.1145/3450337.3483486}. Our research here contributes to this field of study by introducing a novel method for predicting Hexad types based on people's gamer archetypes.

\section{Method}
This section describes the details of our participant recruitment, the survey collection, Stack Exchange data, correlation analysis, and the final ablation study. 
\subsection{Participant Recruitment}
To increase the diversity of our research participants, we used a variety of sources to recruit them, including Stack Exchange chat rooms and social media platforms such as Facebook and Twitter.\footnote{Two moderators from Stack Exchange assisted us throughout the recruitment process.} Inclusion criteria were 1) age over 18 years; 2) more than \textit{six months} of active participation in at least one Stack Exchange forum;\footnote{This measure allows us to avoid studying the behavior of churned users \cite{10.1145/3555553}.} and 3) having at least some \textit{occasional} engagement with games and different gamification schemes (self-reported via seven-point Likert scales).\footnote{i.e., 3/7 on the Likert scale, where 7/7 points out the maximum engagement.}
 
In this way, we recruited 67 Stack Exchange users (23 female) with different demographics and varying levels of exposure to games and gamification. The age range of the participants is between 18 and 45 years (mean = 28.21, SD = 9.36). The ethnic distribution of our research participants is as follows: Asian (n = 27), White (n = 14), Hispanic and Latino/a (n = 10), Black or African American (n = 7), Native American (n = 2), mixed race (n = 2), and other races (n = 5). According to our initial screenings, on the seven-point Likert scale, the average user engagements of our participants with games and various gamification schemes (prior to our research) are 4.82 (SD = 1.12) and 4.30 (SD = 0.93), respectively.\footnote{Again, 7/7 shows the maximum engagement.}

\subsection{Survey Collection}
All of our research procedures have been approved by the IRB at the Hong Kong University of Science and Technology. After obtaining informed consent from all research participants, we asked them (via email) to complete four online surveys (three for gamer archetypes and one for Hexad) within one to two days of receiving the email. The HKUST Qualtrics website was used to collect all responses, allowing participants to take breaks between (and during) surveys and to measure the time it took them to complete each survey. Upon successfully receiving the answers, each participant was given HK\$ 40 as a token of appreciation. All survey responses for each participant were stored under their unique Stack Exchange IDs for further analysis. The average completion times (in minutes) for the Bartle, Big Five, and BrainHex are 22.83 (SD = 7.38), 15.25 (SD = 6.94), and 29.50 (SD = 5.66). The average time to complete the Hexad survey is 20.25 minutes (SD = 4.18).
\subsection{Stack Exchange Data}
Stack Exchange Data Explorer\footnote{https://data.stackexchange.com/} was used to collect each participant's CQA data using their Stack Exchange IDs. This data includes information about each participant's posting behavior, i.e., their posts' exact timestamps and peer-evaluated posting quality scores.\footnote{The difference between the number of \textit{up-votes} and \textit{down-votes}} Different types of posts in the CQA of Stack Exchange include \textit{questions}, \textit{answers}, and \textit{comments}. In this context, user posts \textit{closed} and \textit{removed} by community moderators are regarded as \textit{spam}. \textit{Closed} posts with \textit{negative scores} are also considered \textit{spam}. 

In addition to users' posting data, we also collected their metadata on other types of activities, i.e., their voting behavior and edit services. 

The oldest post in our dataset was made on July 10th, 2015, and the most recent post was made on September 18th, 2022. The average reputation of research participants in this work is 9,656.04 (SD = 10,067.18). And the number of users in the categories of novice users, low reputation users, established users, and trusted users are n = 8, n = 17, n = 23, and n = 19 respectively. The average number of questions, answers, comments, and spam posts per individual is 43.76 (SD = 26.12), 52.10 (SD = 29.18), 68.11 (SD = 35.92), and 4.08 (SD = 2.23), respectively. Furthermore, the average number of votes and edits per person is 496.92 (SD = 311.55) and 30.74 (SD = 26.62), respectively. The most dominant behavior (action) of each participant in our study is determined by counting the frequency with which user behaviors are repeated.
\subsection{Correlation Analysis and Ablation Study}
In keeping with the exploratory spirit of our study, we first conduct and present the results from a thorough Pearson correlation analysis on data from collected \textit{game} and \textit{gamification} surveys and data from the participants' \textit{CQA activities}.

Following that, we train the computational models for our prediction tasks (i.e., predicting the dominant user behavior and Hexad type). To this end, we employ a high-performance machine learning model known as Extreme Gradient Boosting (which is also called XGBoost). The reason for using XGBoost is its simplicity and strength in reducing the training bias \cite{10.1145/3449086}. We build and compare our models based on various input features derived from our participants' previous survey responses (i.e., Bartle, Big Five, and BrainHex characters). We compare our models (also) against the baseline of a random labeler. The reason for using a random labeler here is because, to the best of our knowledge, there are currently no other (suitable/better) baselines that can be used for comparison with our work. Due to the limited amount of our sample data (for training purposes), we use the metric of five-fold cross-validation error to evaluate and report the performance of all models in this paper. After dividing our data into different folds (for the purpose of cross-validation), we use an up-sampling technique (based on \textit{Bootstrapping}) to address the issue of imbalanced data labels. All models in our study are created and tested on a local PC equipped with an Intel i7-5820k CPU (6 cores @3.3GHz) and 32GB RAM using the popular Python library \textit{scikit-learn}.
\begin{figure}[!t]
\centering
    \subfloat[BrainHex and Questions]{
    \includegraphics[width=0.49\columnwidth]{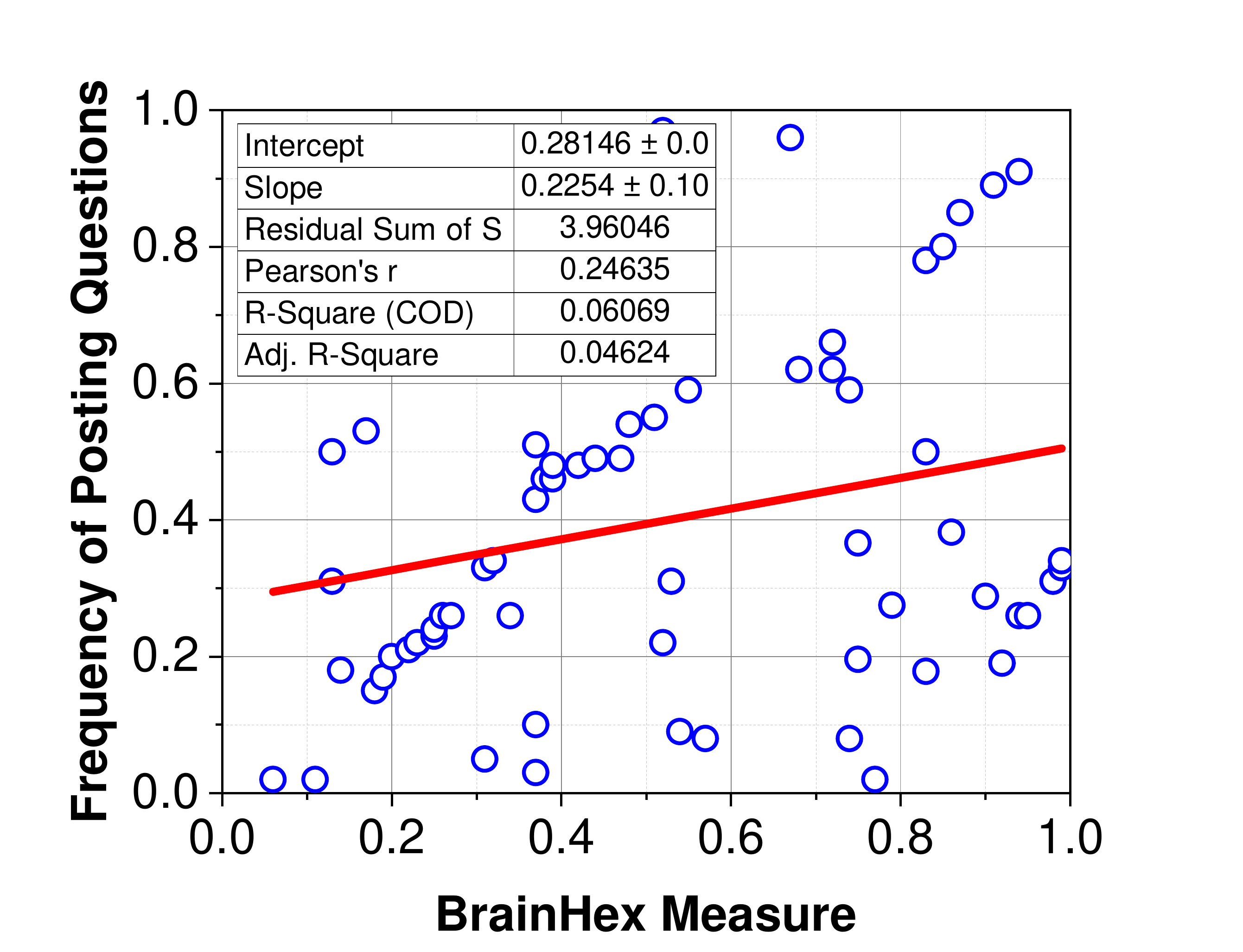}
  \label{fig:corr1}
  }
  \subfloat[BrainHex and Answers]{
  \includegraphics[width=0.49\columnwidth]{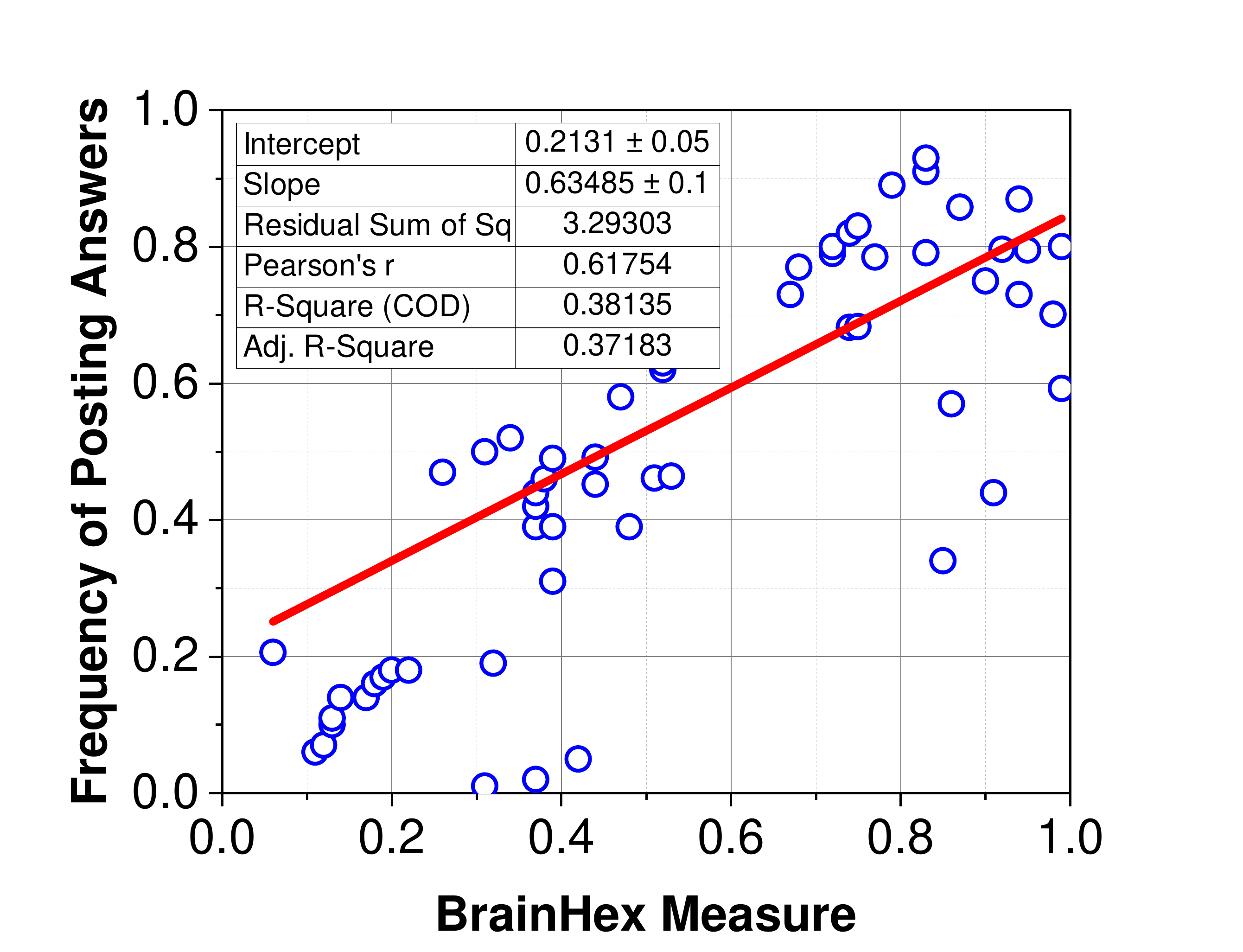}
  \label{fig:corr2}
  }
\\
    \subfloat[BrainHex and Comments]{
  \includegraphics[width=0.49\columnwidth]{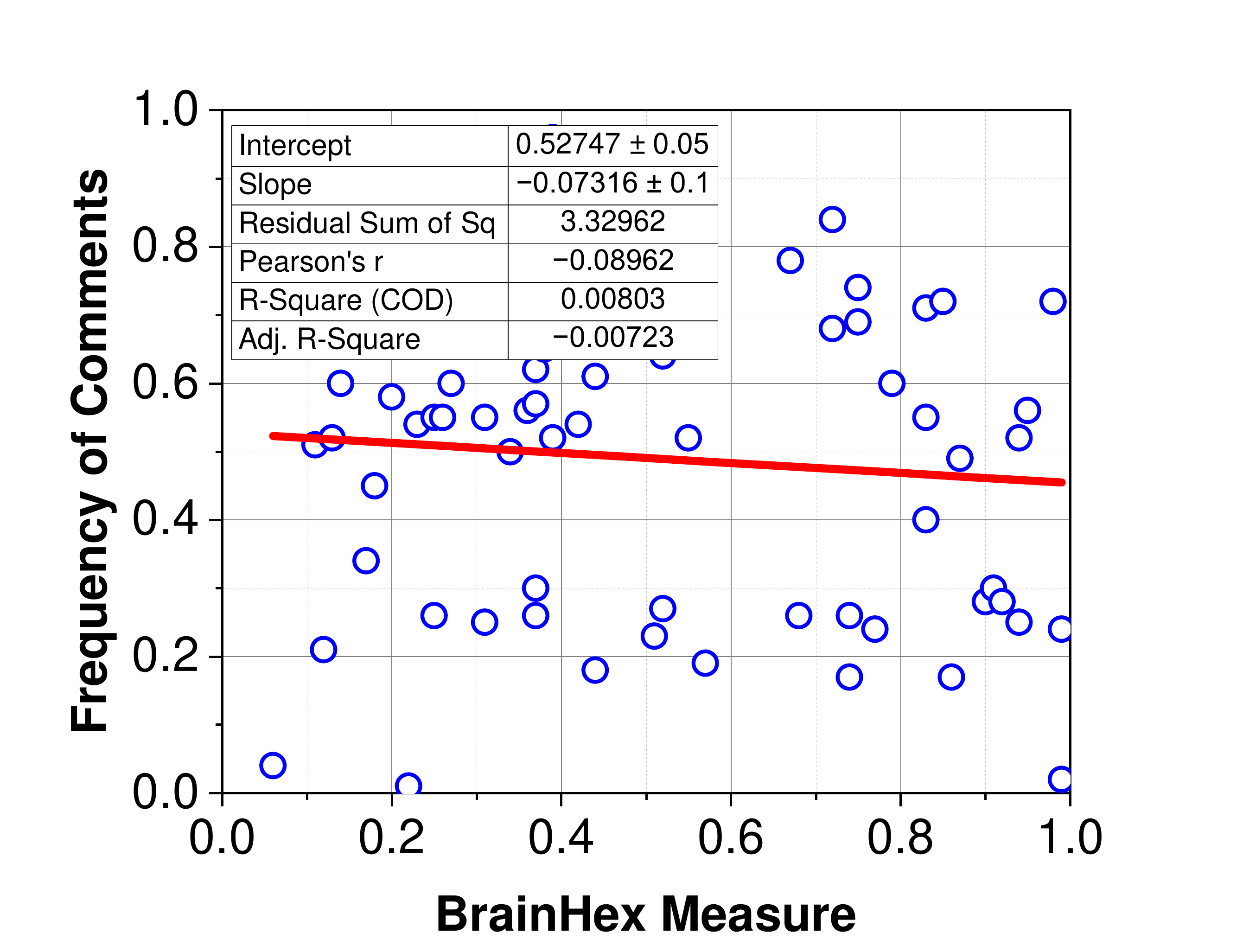}
  \label{fig:corr3}
  }
  \subfloat[BrainHex and Spams]{
  \includegraphics[width=0.49\columnwidth]{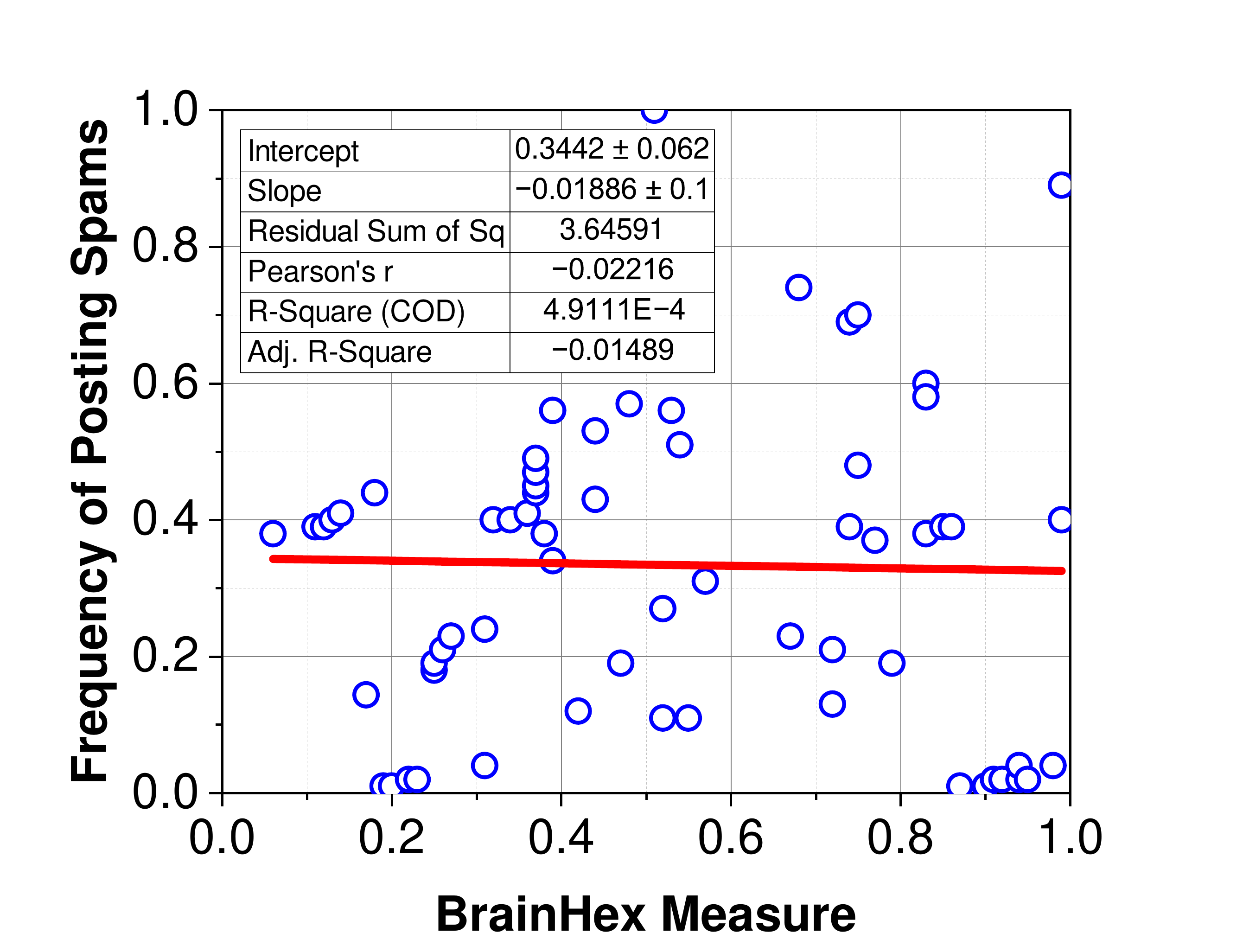}
  \label{fig:corr4}
  }
\caption{The correlation plots between seeker participants' BrainHex measures (intensities) and the frequencies of their (a) questions, (b) answers, (c) comments, and (d) spam posts.}
\label{fig:corrplots}
\end{figure}
\section{Findings}
This section presents the findings of our study in three parts: 1) a report of interesting correlations, 2) a report of dominant user behavior prediction, and 3) a report of Hexad user type prediction.
\begin{figure}[!t]
    \subfloat[BrainHex and Philanthropist Measures]{
  	\includegraphics[width=0.49\columnwidth]{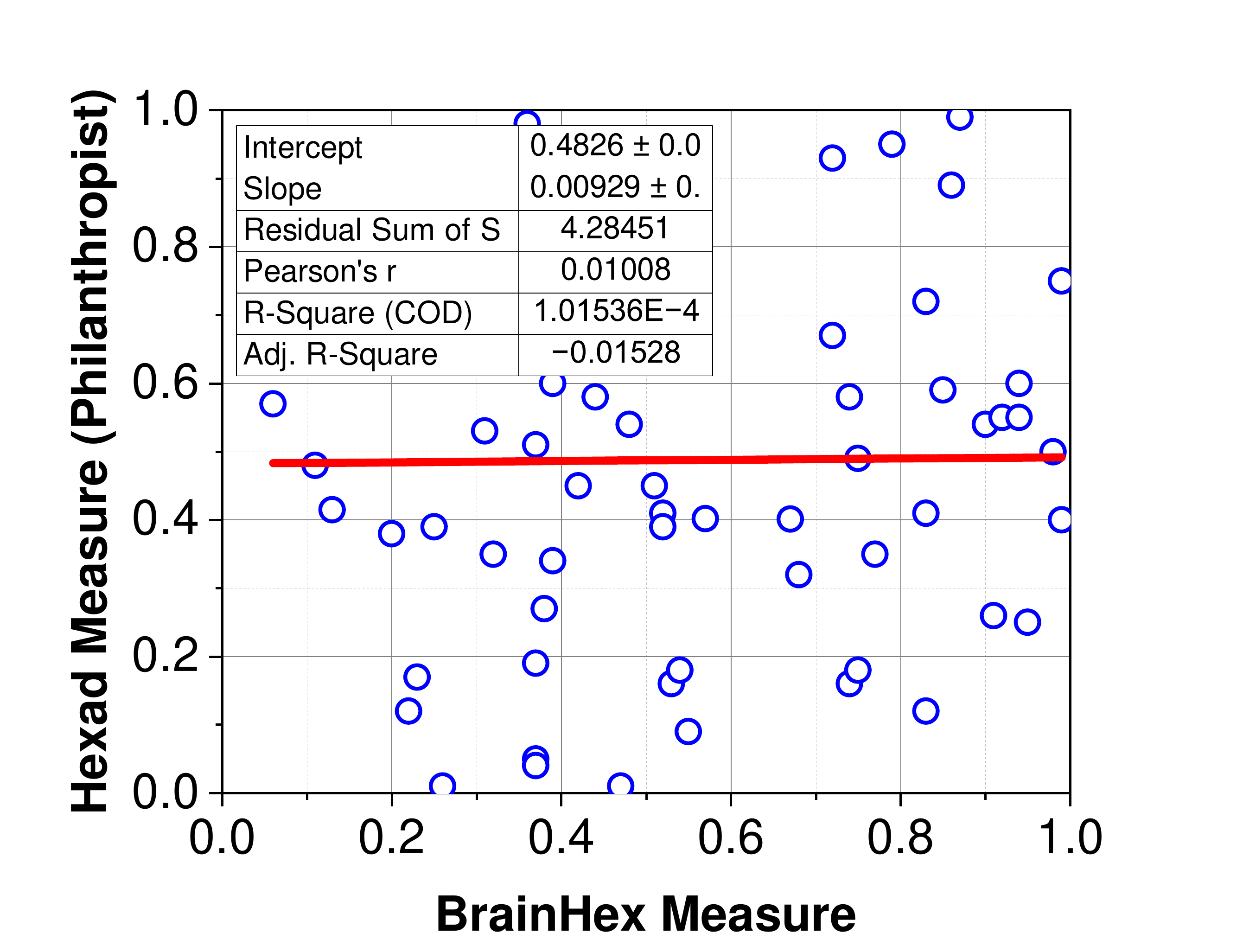}
  \label{fig:hx1}
  }
  \subfloat[BrainHex and Socializer Measures]{
  	\includegraphics[width=0.49\columnwidth]{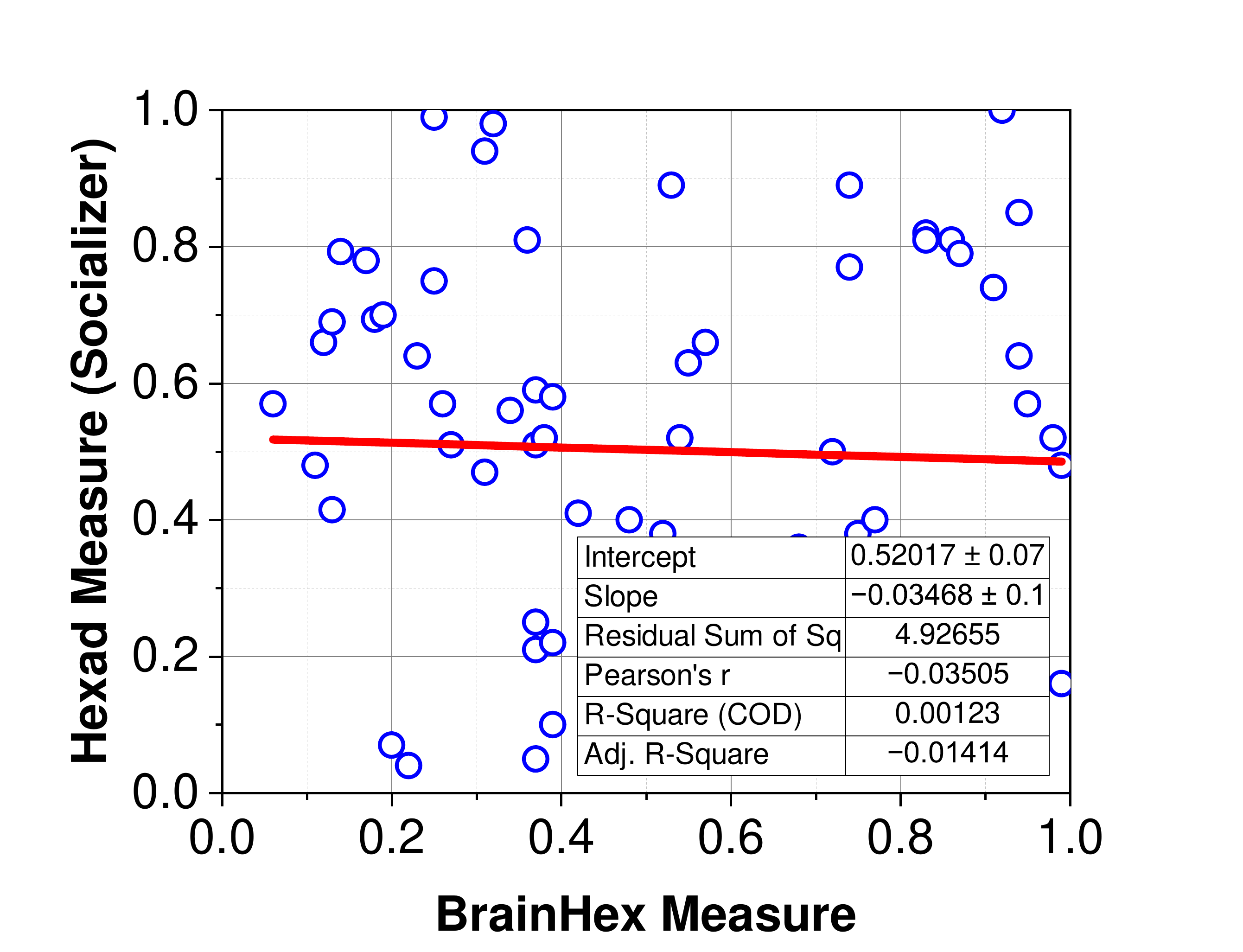}
  \label{fig:hx2}
  }
  \\
    \subfloat[BrainHex and Free Spirit Measures]{
  	\includegraphics[width=0.49\columnwidth]{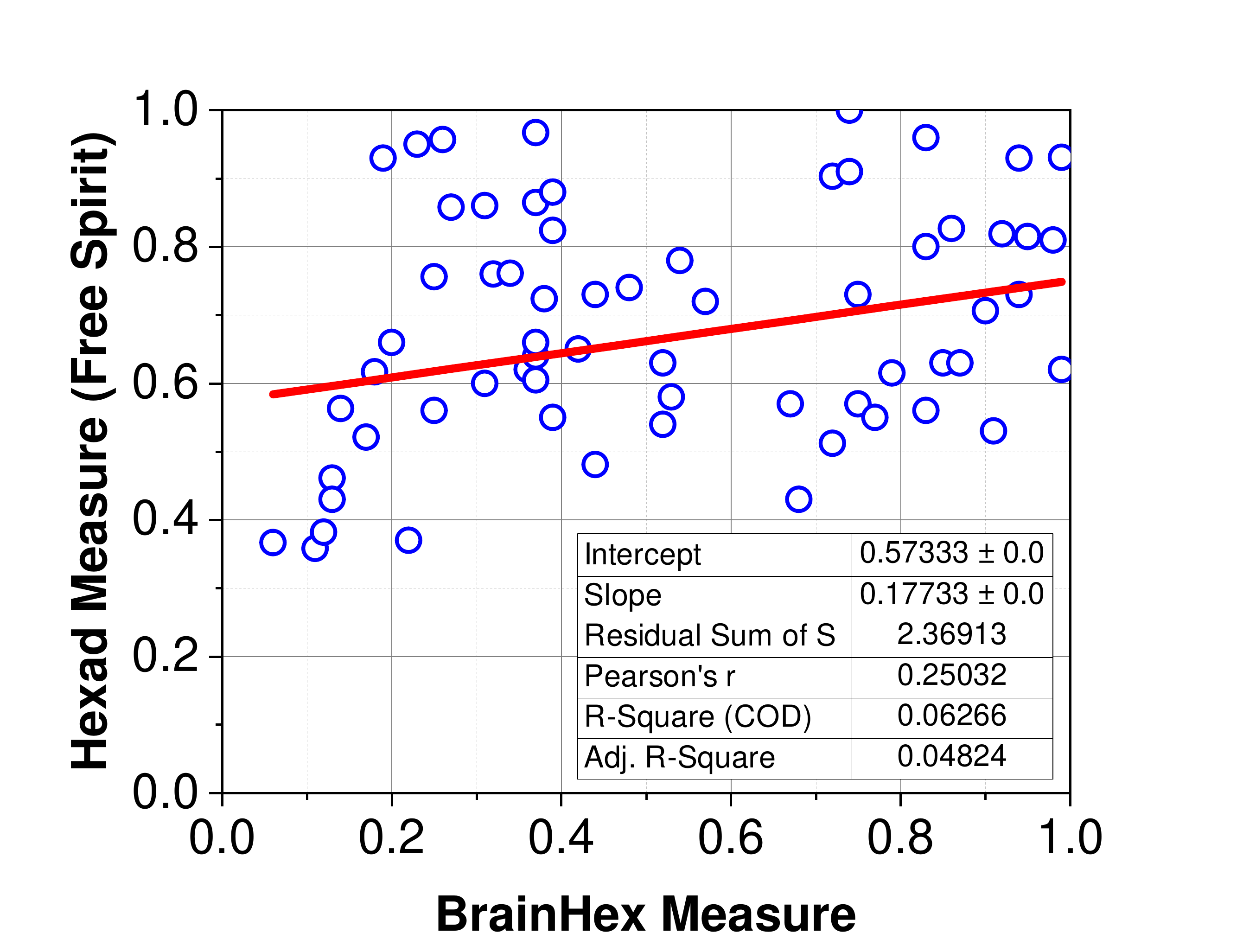}
  \label{fig:hx3}
  }
    \subfloat[BrainHex and Achiever Measures]{
  	\includegraphics[width=0.49\columnwidth]{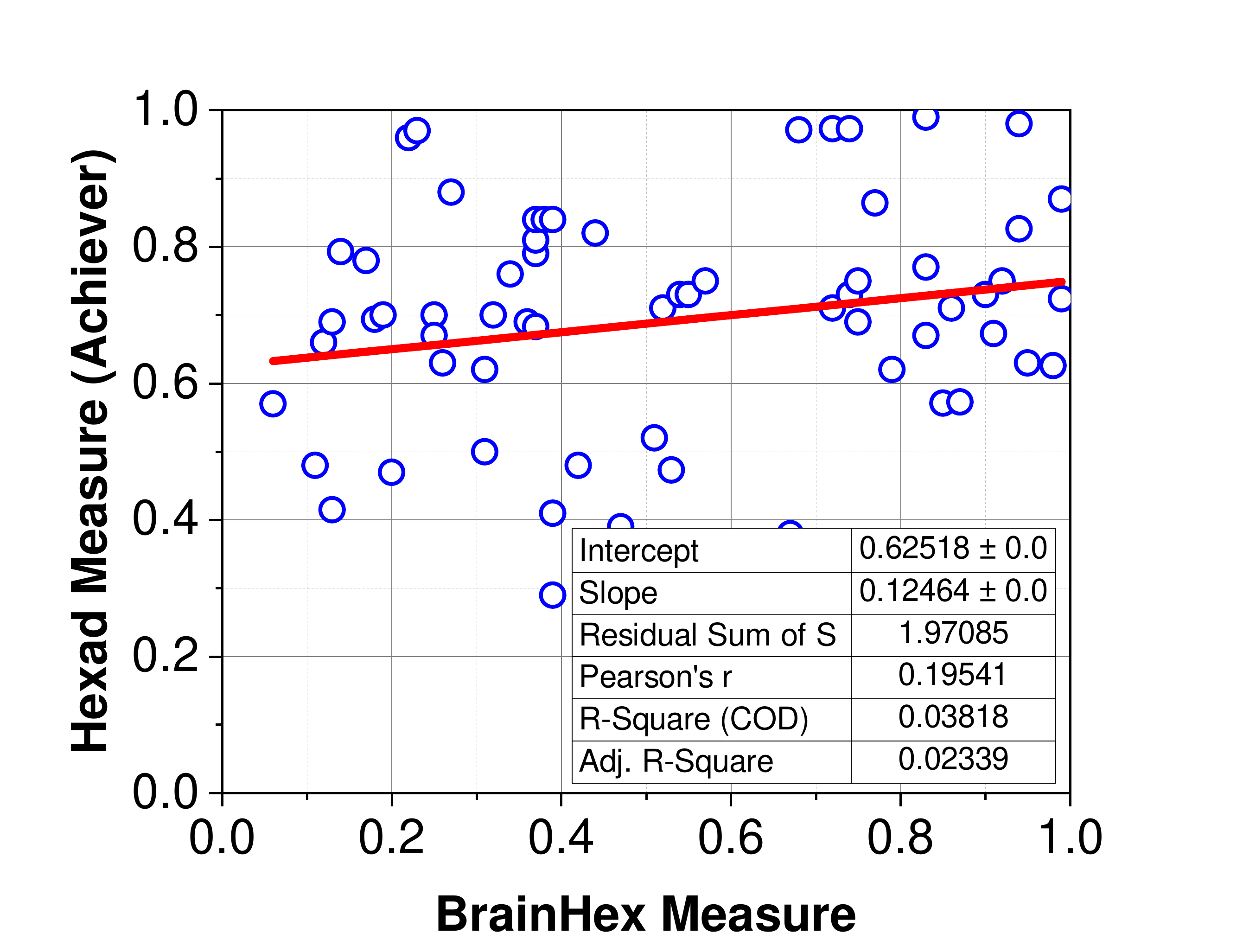}
  \label{fig:hx4}
  }
  \\
    \subfloat[BrainHex and Player Measures]{
  	\includegraphics[width=0.49\columnwidth]{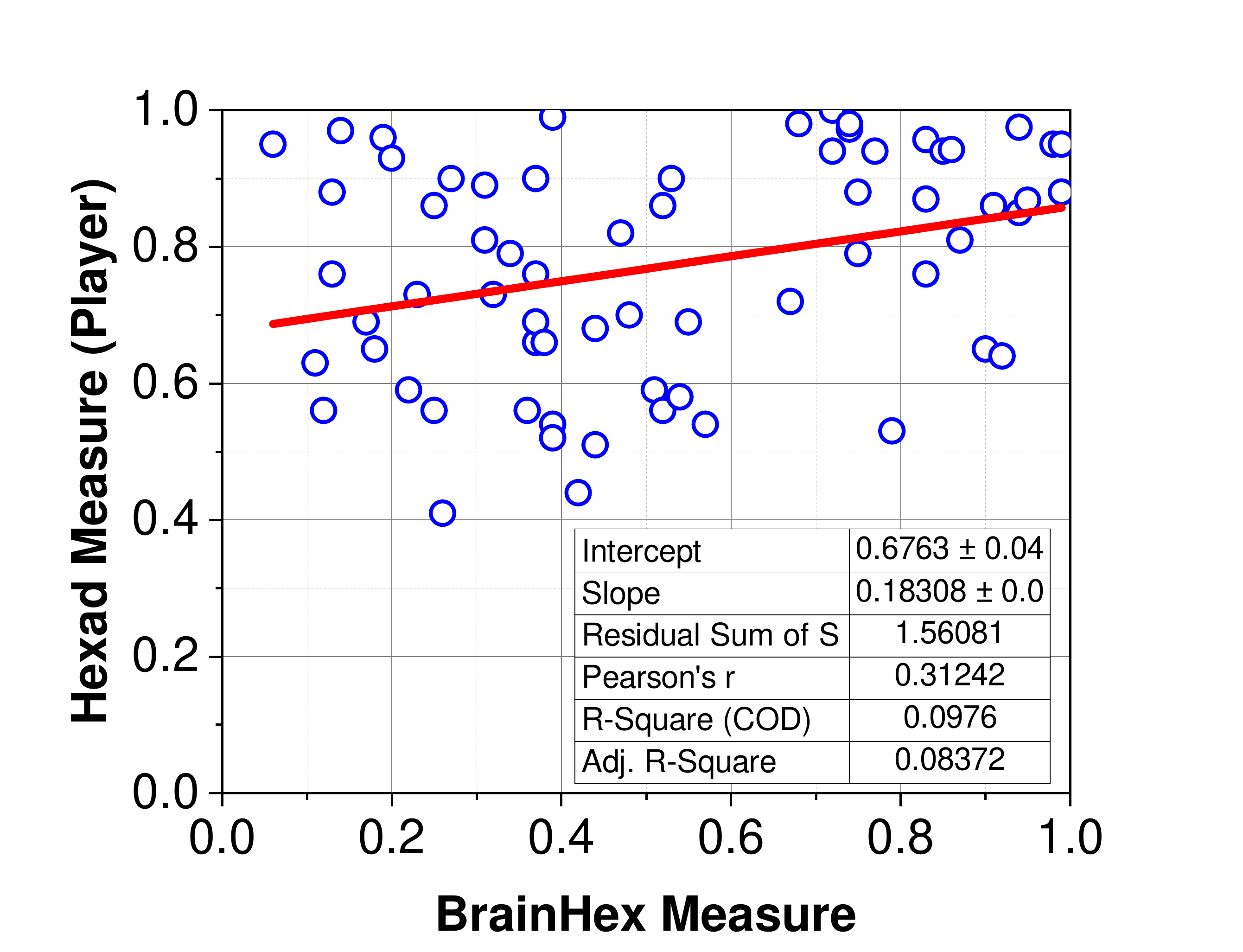}
  \label{fig:hx5}
  }
    \subfloat[BrainHex and Disruptor Measures]{
  	\includegraphics[width=0.49\columnwidth]{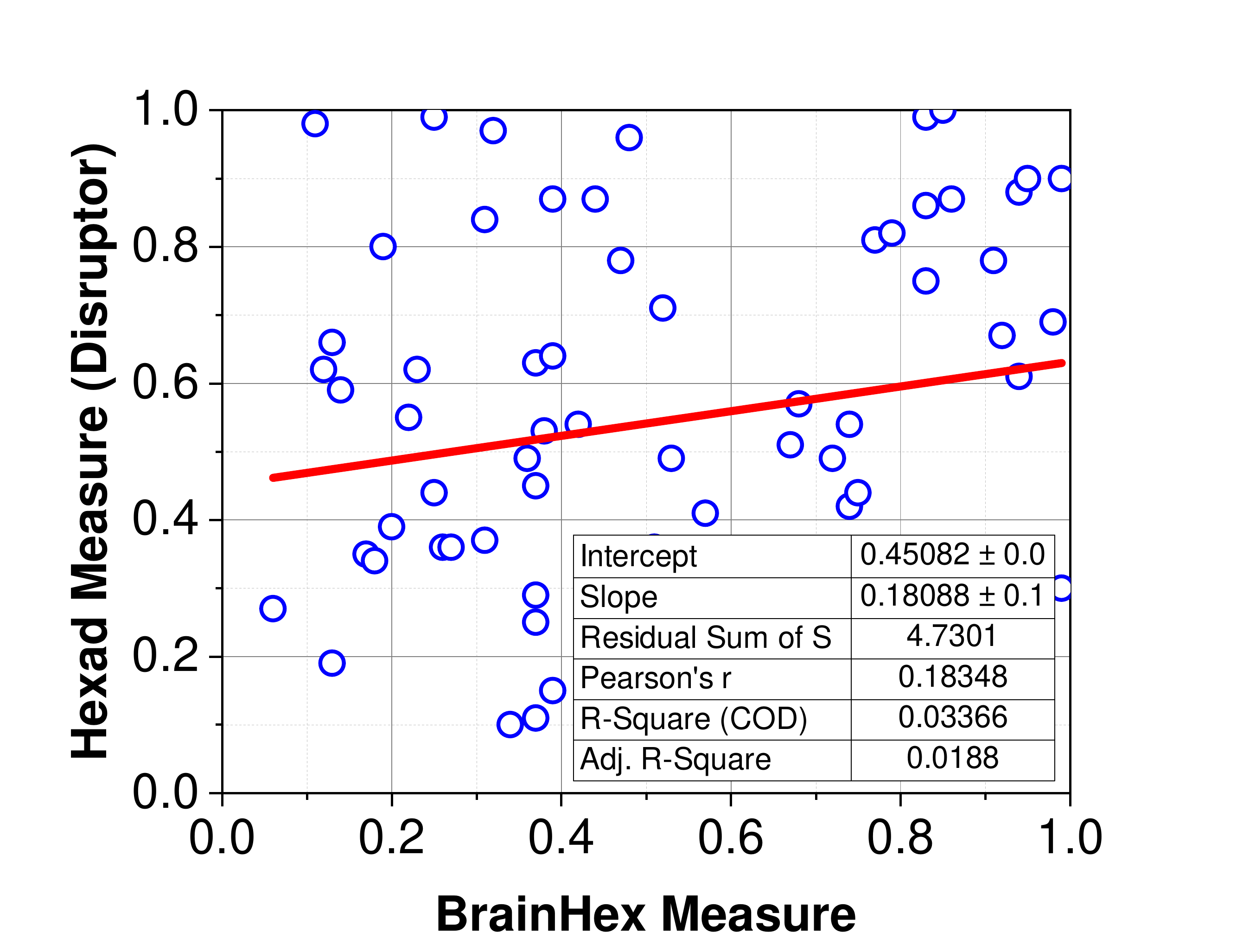}
  \label{fig:hx6}
  }
\caption{The correlation plots between seeker participants' BrainHex and Hexad type measures (intensities).}
\label{fig:corrHex}
\end{figure}
\subsection{Interesting Correlations}
We conducted n = 192 Pearson correlation tests to uncover the interesting relationships between participants' game-based characters and behavioral profiles in the CQA and between participants' game-based characters and gamification-based characters in general (i.e., Hexad type). For the purpose of illustration, Figure \ref{fig:corrplots} depicts the correlation plots (tested with a statistical confidence of 99.95) between seeker participants' BrainHex measures (intensities) and the frequencies of seeker users' posting behaviors, i.e., questions (Pearson's r value = + 0.24), answers (Pearson's r value = + 0.61), comments (Pearson's r value = - 0.08), and spam posts (Pearson's r value = - 0.02). We observe a statistically significant positive correlation between the seekers' BrainHex measures and their tendencies to post (more) questions and answers inside the CQA community. However, the statistical significance of the negative correlation between seekers' BrainHex measures and their frequencies of commenting and spamming still needs to be proved.

Likewise, Figure \ref{fig:corrHex} shows the correlation plots (tested with a statistical confidence of 99.95) between seeker participants' BrainHex and Hexad type measures (intensities). The Pearson r values resolved for the Hexad type correlations are as follows: Philanthropist (r = + 0.01), Socializer (r = - 0.03), Free Spirit (r = + 0.25), Achiever (r = + 0.19), Player (r = + 0.31), and Disruptor (r = + 0.18). In the case of the seeker archetype (from BrainHex), only the positive correlations reported for \textit{free spirit} and \textit{player} Hexad measures are statistically significant. 

It is worth noting here that different surveys often yield different scales for indicating user characteristics and that there is often a considerable discrepancy between the frequency with which users engage in various activities. With this point in mind, we have therefore normalized all numbers in all figures (and analyses) to lie within the range [0,1] for greater clarity and better comparability.
\begin{table}[t!]
\centering
\caption{The average cross-validation (CV) error performance of the various prediction models}~\label{tab:pres1}
\includegraphics[scale=0.68]{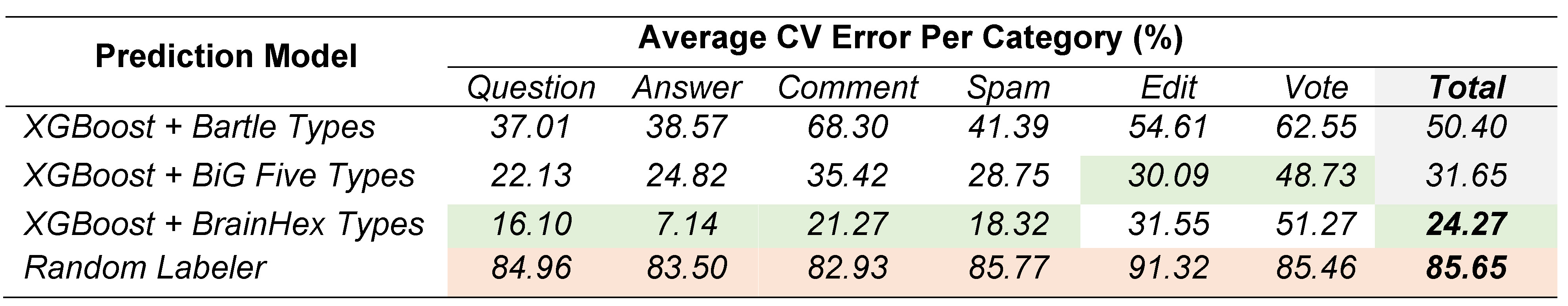}
\end{table}
The following are the most important statistically significant findings and conclusions from our correlation analysis:

\textbf{Bartle.} The frequency of questions and answers in the CQA correlates positively with Bartle's \textit{achiever} and \textit{Socializer} measures. We found that the amount of comments and spam posted in the CQA is positively correlated with Bartle's \textit{explorer} and \textit{killer} measures. Bartle's \textit{achiever} measure correlates positively with Hexad's \textit{free spirit}, \textit{achiever}, and \textit{player} intensities. There is also a positive correlation between Bartle's \textit{killer} measure and Hexad's \textit{achiever}, \textit{player}, and \textit{disruptor} metrics (indicators).

\textbf{Big Five.} The number of questions asked, answers given, comments written, and edits made in the CQA correlate positively with the Big Five measures of \textit{openness} and \textit{extraversion}. One of the Big Five dimensions, \textit{conscientiousness}, correlates positively with the frequencies of voting and editing, whereas \textit{neuroticism} correlates negatively with both. Furthermore, the Hexad measures \textit{socializer} and \textit{philanthropist} correlate positively with the Big Five levels of \textit{extraversion} and \textit{agreeableness}. Also, there appears to be a positive relationship between \textit{openness} and Hexad's \textit{free spirit} measure.

\textbf{BrainHex.} The frequency of CQA posts (of any type) correlates positively with the BrainHex \textit{mastermind} and \textit{conqueror} measures. The frequency of questions and answers is proportional to the BainHex \textit{seeker} measure. The BrainHex \textit{daredevil} metric is positively correlated with the frequency of votes and spam. The \textit{socializer} measure, on the other hand, is positively correlated with the frequency of comments and answers. The number of questions a participant asks correlates positively with their BrainHex \textit{survival} measure, while the number of answers correlates negatively with it. Nonetheless, the BrainHex \textit{achiever} measurement is positively correlated with the number of answers provided by the participants identified with that character. In addition, the BrainHex \textit{achiever} metric is positively correlated with Hexad's \textit{philanthropist}, \textit{achiever}, and \textit{player} intensities. The seeker archetype in BrainHex, as noted before, is positively correlated to the measurements of \textit{free spirit} and \textit{player} archetypes in Hexad. In BrainHex, the \textit{daredevil} metric is positively correlated with the \textit{player} and \textit{disruptor} measurements in Hexad. The \textit{socializer} measure in the BrainHex is positively related to the \textit{philanthropist} and \textit{socializer} measures in Hexad.

\subsection{Predicting the Dominant User Behavior}
Table \ref{tab:pres1} summarizes the performances of the different prediction models based on their average cross-validation error. Overall, the \textbf{combination of XGBoost and the BrainHex user type measures} outperforms all other competing models in predicting the most dominant user behavior. This combination outperforms the Big Five and Bartle models by 26.13\% and 7.38\%, respectively. In addition, BrainHex consistently outperforms all other models at the categorical level, with the exception of the \textit{editing} and \textit{voting} tasks (where the BrainHex model falls slightly behind the Big Five model). Furthermore, we can see that all gamer archetypes can significantly outperform the prediction of a random labeler. Hence, even in the absence of BrainHex data, the Big Five and then the Bartle model could be used as helpful alternatives. However, the accuracy issue should be considered for more sensitive tasks and decisions.

\subsection{Predicting the Hexad Type}
Similarly, Table \ref{tab:pers2} summarizes the performances of the various prediction models when predicting the (most dominant) Hexad type. Again, the \textbf{combination of XGBoost and the BrainHex user type measures} outperforms all other competing models in accurately predicting the Hexad types. According to the \textit{total} performance results, this model (XGBoost + BrainHex) reduces error by 15.60\% compared to the Bartle archetype and 3.77\% compared to the Big Five model. Furthermore, with the exception of the \textit{philanthropist} user type (where BrainHex falls 1.85\% short of the Big Five model), the BrainHex model (also) outperforms all other competing models at all categorical levels. We should mention here that all gamer archetypes (i.e., Bartle, Big Five, and BrainHex) can significantly outperform the prediction of a random labeler. Hence, in the absence of BrainHex data, the Big Five and then the Bartle model could still be used as viable (helpful) substitutes. 

\begin{table}[t!]
\centering
\caption{The performance of various models in predicting the Hexad type of the participants}~\label{tab:pers2}
\includegraphics[scale=0.58]{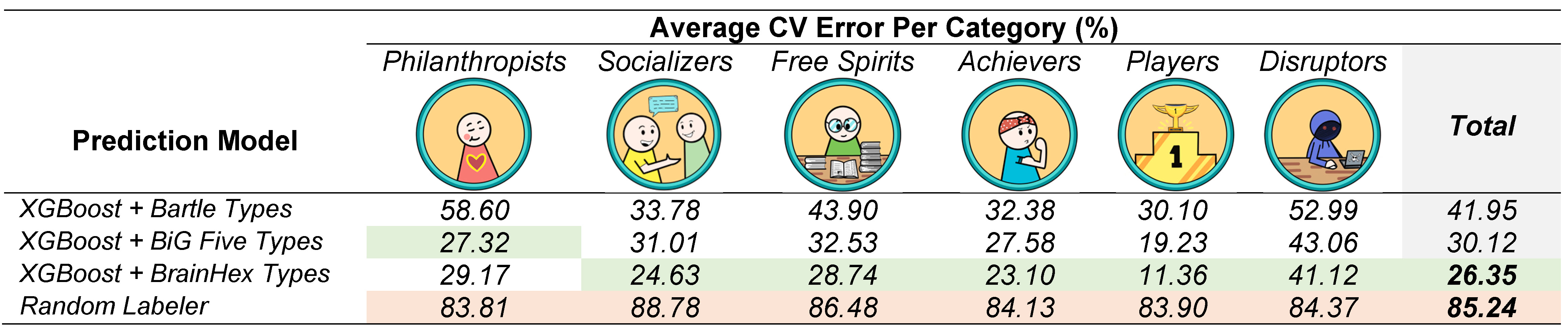}
\end{table}

\section{Discussion}
In this section, we will discuss some of the major limitations of our research and set an agenda for future work.
\subsection{Limitations}
There are several critical limitations to our research in this paper that should be mentioned. The first and most obvious limitation is the small number of participants in the study. Unfortunately, it is becoming increasingly difficult to recruit participants for research studies such as ours that require lengthy and time-consuming surveys to be completed. Also, due to the limited number of data samples, we decided not to use Deep Learning or other advanced neural network models for our prediction tasks. Future work could focus on collecting more data (on a larger scale) and improving the machine-learning component of the models. In addition to these caveats, the context of our research study itself is a major limitation. Obviously, Stack Exchange is not the best representative for all kinds of online gamified platforms on the Internet. Hence, it is necessary that future studies cross-check the findings of our work also on other (and more diverse) platforms.

\subsection{Agenda for Future Research}
In the end, we present four interesting research directions for future work. 1) Future player surveys and gamification archetypes (or data collection methods) will need to be less intrusive and collect as little personal information as possible. In addition, the advent of platforms such as the metaverse has necessitated the development of newer and more secure techniques for collecting and disseminating human data. As an idea for future research, we would like to see if it is possible to use people's metaverse-based peripheral data to infer aspects of their personality in order to provide them with better and more tailored gaming/gamification services. 2) Understanding how users and gamification experts perceive and reason about the relationships between different gamer archetypes, CQA behaviors, and Hexad is another interesting research direction for future work that merits being thoroughly explored. Comprehensive qualitative research methods, such as inductive and deductive content analysis, are likely to yield better results in this type of investigation. 3) People's interest in the playful aspects of a gamified platform may be a good indicator of their willingness to stay on that site for a more extended period of time. Future research could examine and explore this idea further to determine if information about users' gamer and gamification archetypes could help gamified platforms and services detect when a user will churn their websites/services (or become less active/loyal), perhaps in a more timely and accurate manner. 4) Although we found some interesting correlations between gamer archetypes, user behaviors, and Hexad types, our study does not demonstrate or assert any causal relationships. Future research could use controlled experimental designs to determine the presence (or absence) of any causal relationships.


\section{Conclusion}
In this paper, we investigated whether information about users' game-related data (collected through a series of gamer-archetype surveys) could be used to predict their most dominant behaviors and Hexad user types in a non-gaming (but still gamified) environment. We found some promising results: users' game-related data (i.e., Bartle, Big Five, and BrainHex) could all provide valuable information about their behavioral characteristics and Hexad user types. However, \textit{BrainHex} outperforms the other gaming archetypes studied in this work on the prediction tasks. Finally, we presented our research agenda for future work.

\section{Acknowledgement}
The authors would like to thank Mr. Rahman Hadi Mogavi for proofreading and contributing beautiful images to this paper. This research has been supported in part by the MetaHKUST project from the Hong Kong University of Science and Technology (Guangzhou), and 5GEAR and FIT projects from the Academy of Finland.
\bibliographystyle{splncs04}
\bibliography{myref.bib}

\begin{thebibliography}{10}
\providecommand{\url}[1]{\texttt{#1}}
\providecommand{\urlprefix}{URL }
\providecommand{\doi}[1]{https://doi.org/#1}

\bibitem{10.1145/3341215.3356266}
Altmeyer, M., Lessel, P., Schubhan, M., Kr\"{u}ger, A.: Towards predicting
  hexad user types from smartphone data. In: Extended Abstracts of the Annual
  Symposium on Computer-Human Interaction in Play Companion Extended Abstracts.
  p. 315–322. CHI PLAY '19 Extended Abstracts, Association for Computing
  Machinery, New York, NY, USA (2019). \doi{10.1145/3341215.3356266},
  \url{https://doi.org/10.1145/3341215.3356266}

\bibitem{10.1145/3410404.3414232}
Altmeyer, M., Tondello, G.F., Kr\"{u}ger, A., Nacke, L.E.: Hexarcade:
  Predicting hexad user types by using gameful applications. In: Proceedings of
  the Annual Symposium on Computer-Human Interaction in Play. p. 219–230. CHI
  PLAY '20, Association for Computing Machinery, New York, NY, USA (2020).
  \doi{10.1145/3410404.3414232}, \url{https://doi.org/10.1145/3410404.3414232}

\bibitem{10.1145/2488388.2488398}
Anderson, A., Huttenlocher, D., Kleinberg, J., Leskovec, J.: Steering user
  behavior with badges. In: Proceedings of the 22nd International Conference on
  World Wide Web. p. 95–106. WWW '13, Association for Computing Machinery,
  New York, NY, USA (2013). \doi{10.1145/2488388.2488398},
  \url{https://doi.org/10.1145/2488388.2488398}

\bibitem{Barr2021}
Barr, M., Copeland-Stewart, A.: Playing video games during the {COVID}-19
  pandemic and effects on players' well-being. Games and Culture
  \textbf{17}(1),  122--139 (May 2021). \doi{10.1177/15554120211017036},
  \url{https://doi.org/10.1177/15554120211017036}

\bibitem{bartle1996hearts}
Bartle, R.: Hearts, clubs, diamonds, spades: Players who suit muds. Journal of
  MUD research  \textbf{1}(1), ~19 (1996)

\bibitem{bateman2011player}
Bateman, C., Lowenhaupt, R., Nacke, L.E., et~al.: Player typology in theory and
  practice. In: DiGRA conference. pp. 1--24. Citeseer (2011)

\bibitem{Boggio2020}
Boggio, C., Moscarola, F.C., Gallice, A.: What is good for the goose is good
  for the gander? Economics of Education Review  \textbf{75},  101952 (Apr
  2020). \doi{10.1016/j.econedurev.2019.101952},
  \url{https://doi.org/10.1016/j.econedurev.2019.101952}

\bibitem{Cairns2019}
Cairns, P., Power, C., Barlet, M., Haynes, G.: Future design of accessibility
  in games: A design vocabulary. International Journal of Human-Computer
  Studies  \textbf{131},  64--71 (Nov 2019). \doi{10.1016/j.ijhcs.2019.06.010},
  \url{https://doi.org/10.1016/j.ijhcs.2019.06.010}

\bibitem{Chesham2017}
Chesham, A., Wyss, P., M\"{u}ri, R.M., Mosimann, U.P., Nef, T.: What older
  people like to play: Genre preferences and acceptance of casual games. {JMIR}
  Serious Games  \textbf{5}(2), ~e8 (Apr 2017). \doi{10.2196/games.7025},
  \url{https://doi.org/10.2196/games.7025}

\bibitem{Cole2007}
Cole, H., Griffiths, M.D.: Social interactions in massively multiplayer online
  role-playing gamers. CyberPsychology $\&$ Behavior  \textbf{10}(4),  575--583
  (Aug 2007). \doi{10.1089/cpb.2007.9988},
  \url{https://doi.org/10.1089/cpb.2007.9988}

\bibitem{Connolly2012}
Connolly, T.M., Boyle, E.A., MacArthur, E., Hainey, T., Boyle, J.M.: A
  systematic literature review of empirical evidence on computer games and
  serious games. Computers \& Education  \textbf{59}(2),  661--686 (Sep 2012).
  \doi{10.1016/j.compedu.2012.03.004},
  \url{https://doi.org/10.1016/j.compedu.2012.03.004}

\bibitem{10.1145/2181037.2181040}
Deterding, S., Dixon, D., Khaled, R., Nacke, L.: From game design elements to
  gamefulness: Defining "gamification". In: Proceedings of the 15th
  International Academic MindTrek Conference: Envisioning Future Media
  Environments. p. 9–15. MindTrek '11, Association for Computing Machinery,
  New York, NY, USA (2011). \doi{10.1145/2181037.2181040},
  \url{https://doi.org/10.1145/2181037.2181040}

\bibitem{ESA_Survey2022}
(ESA), T.E.S.A.: Essential facts about the video game industry (November 2022),
  \url{https://www.theesa.com/resource/2022-essential-facts-about-the-video-game-industry/}

\bibitem{Galleguillos2019}
Galleguillos, L., Santelices, I., Bustos, R.: Designing a board game for
  industrial engineering students. a collaborative work experience of freshmen.
  In: {INTED}2019 Proceedings. {IATED} (Mar 2019).
  \doi{10.21125/inted.2019.0075},
  \url{https://doi.org/10.21125/inted.2019.0075}

\bibitem{Gosling2003}
Gosling, S.D., Rentfrow, P.J., Swann, W.B.: A very brief measure of the
  big-five personality domains. Journal of Research in Personality
  \textbf{37}(6),  504--528 (Dec 2003). \doi{10.1016/s0092-6566(03)00046-1},
  \url{https://doi.org/10.1016/s0092-6566(03)00046-1}

\bibitem{Granic2014}
Granic, I., Lobel, A., Engels, R.C.M.E.: The benefits of playing video games.
  American Psychologist  \textbf{69}(1),  66--78 (Jan 2014).
  \doi{10.1037/a0034857}, \url{https://doi.org/10.1037/a0034857}

\bibitem{BritWiki}
Grice, J.W., Doorey, M., Lotha, G.: Five-factor model of personality (January
  2019),
  \url{https://www.britannica.com/science/five-factor-model-of-personality}

\bibitem{Haberlin2022}
Haberlin, K.A., Atkin, D.J.: Mobile gaming and internet addiction: When is
  playing no longer just fun and games? Computers in Human Behavior
  \textbf{126},  106989 (Jan 2022). \doi{10.1016/j.chb.2021.106989},
  \url{https://doi.org/10.1016/j.chb.2021.106989}

\bibitem{10.1145/3491140.3528274}
Hadi~Mogavi, R., Guo, B., Zhang, Y., Haq, E.U., Hui, P., Ma, X.: When
  gamification spoils your learning: A qualitative case study of gamification
  misuse in a language-learning app. In: Proceedings of the Ninth ACM
  Conference on Learning @ Scale. p. 175–188. L@S '22, Association for
  Computing Machinery, New York, NY, USA (2022). \doi{10.1145/3491140.3528274},
  \url{https://doi.org/10.1145/3491140.3528274}

\bibitem{10.1145/3555553}
Hadi~Mogavi, R., Haq, E.U., Gujar, S., Hui, P., Ma, X.: More gamification is
  not always better: A case study of promotional gamification in a question
  answering website. Proc. ACM Hum.-Comput. Interact.  \textbf{6}(CSCW2) (nov
  2022). \doi{10.1145/3555553}, \url{https://doi.org/10.1145/3555553}

\bibitem{10.1145/3449086}
Hadi~Mogavi, R., Ma, X., Hui, P.: Characterizing student engagement moods for
  dropout prediction in question pool websites. Proc. ACM Hum.-Comput.
  Interact.  \textbf{5}(CSCW1) (apr 2021). \doi{10.1145/3449086},
  \url{https://doi.org/10.1145/3449086}

\bibitem{10.1145/3555124}
Hadi~Mogavi, R., Zhang, Y., Haq, E.U., Wu, Y., Hui, P., Ma, X.: What do users
  think of promotional gamification schemes? a qualitative case study in a
  question answering website. Proc. ACM Hum.-Comput. Interact.
  \textbf{6}(CSCW2) (nov 2022). \doi{10.1145/3555124},
  \url{https://doi.org/10.1145/3555124}

\bibitem{10.1145/3311350.3347167}
Hallifax, S., Serna, A., Marty, J.C., Lavou\'{e}, G., Lavou\'{e}, E.: Factors
  to consider for tailored gamification. In: Proceedings of the Annual
  Symposium on Computer-Human Interaction in Play. p. 559–572. CHI PLAY '19,
  Association for Computing Machinery, New York, NY, USA (2019).
  \doi{10.1145/3311350.3347167}, \url{https://doi.org/10.1145/3311350.3347167}

\bibitem{Hamari2014Work}
Hamari, J., Koivisto, J., Sarsa, H.: Does gamification work? -- a literature
  review of empirical studies on gamification. In: 2014 47th Hawaii
  International Conference on System Sciences. {IEEE} (Jan 2014).
  \doi{10.1109/hicss.2014.377}, \url{https://doi.org/10.1109/hicss.2014.377}

\bibitem{hamari2014player}
Hamari, J., Tuunanen, J.: Player types: A meta-synthesis  (2014)

\bibitem{deHesselle2021}
de~Hesselle, L.C., Rozgonjuk, D., Sindermann, C., Pontes, H.M., Montag, C.: The
  associations between big five personality traits, gaming motives, and
  self-reported time spent gaming. Personality and Individual Differences
  \textbf{171},  110483 (Mar 2021). \doi{10.1016/j.paid.2020.110483},
  \url{https://doi.org/10.1016/j.paid.2020.110483}

\bibitem{10.1145/3450337.3483486}
Kimpen, R., De~Croon, R., Vanden~Abeele, V., Verbert, K.: Towards predicting
  hexad user types from mobile banking data: An expert consensus study. In:
  Extended Abstracts of the 2021 Annual Symposium on Computer-Human Interaction
  in Play. p. 30–36. CHI PLAY '21, Association for Computing Machinery, New
  York, NY, USA (2021). \doi{10.1145/3450337.3483486},
  \url{https://doi.org/10.1145/3450337.3483486}

\bibitem{10.1145/3568732}
Kumar, N., Adams, J.A., Buxton, B., Candy, L., Cesar, P., Clark, L., Cowan,
  B.R., Dey, A., Dugas, P.O.T., Edmonds, E., Goodrich, M.A., Green, M., Grudin,
  J., Kitamura, Y., Konstan, J., Latulipe, C., Lee, M., Malone, T., Mandryk,
  R., Markopoulos, P., Muller, M., Nacke, L., Nakano, Y., Obrist, M.,
  Porcheron, M., Sarcevic, A., Sch\"{o}ning, J., Scott, S., Sharif, B.,
  Steinicke, F., Stumpf, S., Tse, E., Vinayagamoorthy, V.: A chronology of
  sigchi conferences: 1983 to 2022. Interactions  \textbf{29}(6),  34–41 (nov
  2022). \doi{10.1145/3568732}, \url{https://doi.org/10.1145/3568732}

\bibitem{10.1145/3178876.3186147}
Kusmierczyk, T., Gomez-Rodriguez, M.: On the causal effect of badges. In:
  Proceedings of the 2018 World Wide Web Conference. p. 659–668. WWW '18,
  International World Wide Web Conferences Steering Committee, Republic and
  Canton of Geneva, CHE (2018). \doi{10.1145/3178876.3186147},
  \url{https://doi.org/10.1145/3178876.3186147}

\bibitem{paulmetaverse}
Lee, L.H., Braud, T., Zhou, P., Wang, L., Xu, D., Lin, Z., Kumar, A., Bermejo,
  C., Hui, P.: All one needs to know about metaverse: A complete survey on
  technological singularity, virtual ecosystem, and research agenda (2021).
  \doi{10.48550/ARXIV.2110.05352}, \url{https://arxiv.org/abs/2110.05352}

\bibitem{Legaki2020}
Legaki, N.Z., Xi, N., Hamari, J., Karpouzis, K., Assimakopoulos, V.: The effect
  of challenge-based gamification on learning: An experiment in the context of
  statistics education. International Journal of Human-Computer Studies
  \textbf{144},  102496 (Dec 2020). \doi{10.1016/j.ijhcs.2020.102496},
  \url{https://doi.org/10.1016/j.ijhcs.2020.102496}

\bibitem{Lopez2019}
Lopez, C.E., Tucker, C.S.: The effects of player type on performance: A
  gamification case study. Computers in Human Behavior  \textbf{91},  333--345
  (Feb 2019). \doi{10.1016/j.chb.2018.10.005},
  \url{https://doi.org/10.1016/j.chb.2018.10.005}

\bibitem{IHBO}
Ltd, I.H.: Brainhex questionnaire (May 2010),
  \url{http://survey.ihobo.com/BrainHex/}

\bibitem{rezaMogavi2019}
Mogavi, R.H., Gujar, S., Ma, X., Hui, P.: Hrcr: Hidden markov-based
  reinforcement to reduce churn in question answering forums. In: Pacific Rim
  International Conference on Artificial Intelligence, pp. 364--376 (2019)

\bibitem{Nacke2014}
Nacke, L.E., Bateman, C., Mandryk, R.L.: {BrainHex}: A neurobiological gamer
  typology survey. Entertainment Computing  \textbf{5}(1),  55--62 (Jan 2014).
  \doi{10.1016/j.entcom.2013.06.002},
  \url{https://doi.org/10.1016/j.entcom.2013.06.002}

\bibitem{Nicholson2014}
Nicholson, S.: A recipe for meaningful gamification. Gamification in education
  and business pp. 1--20 (2015)

\bibitem{10.1145/2470654.2481341}
Orji, R., Mandryk, R.L., Vassileva, J., Gerling, K.M.: Tailoring persuasive
  health games to gamer type. In: Proceedings of the SIGCHI Conference on Human
  Factors in Computing Systems. p. 2467–2476. CHI '13, Association for
  Computing Machinery, New York, NY, USA (2013). \doi{10.1145/2470654.2481341},
  \url{https://doi.org/10.1145/2470654.2481341}

\bibitem{10.1145/3025453.3025577}
Orji, R., Nacke, L.E., Marco, C.D.: Towards personality-driven persuasive
  health games and gamified systems. In: Proceedings of the 2017 {CHI}
  Conference on Human Factors in Computing Systems. {ACM} (May 2017).
  \doi{10.1145/3025453.3025577}, \url{https://doi.org/10.1145/3025453.3025577}

\bibitem{10.1145/3173574.3174009}
Orji, R., Tondello, G.F., Nacke, L.E.: Personalizing persuasive strategies in
  gameful systems to gamification user types. In: Proceedings of the 2018 CHI
  Conference on Human Factors in Computing Systems. p. 1–14. CHI '18,
  Association for Computing Machinery, New York, NY, USA (2018).
  \doi{10.1145/3173574.3174009}, \url{https://doi.org/10.1145/3173574.3174009}

\bibitem{Parry2014}
Parry, I., Carbullido, C., Kawada, J., Bagley, A., Sen, S., Greenhalgh, D.,
  Palmieri, T.: Keeping up with video game technology: Objective analysis of
  xbox kinect{\texttrademark} and {PlayStation} 3 move{\texttrademark} for use
  in burn rehabilitation. Burns  \textbf{40}(5),  852--859 (Aug 2014).
  \doi{10.1016/j.burns.2013.11.005},
  \url{https://doi.org/10.1016/j.burns.2013.11.005}

\bibitem{reiners2015gami}
Reiners, T., Wood, L.C.: Gami cation in Education and Business. Springer (2015)

\bibitem{Ryan2006}
Ryan, R.M., Rigby, C.S., Przybylski, A.: The motivational pull of video games:
  A self-determination theory approach. Motivation and Emotion  \textbf{30}(4),
   344--360 (Nov 2006). \doi{10.1007/s11031-006-9051-8},
  \url{https://doi.org/10.1007/s11031-006-9051-8}

\bibitem{Seaborn2015}
Seaborn, K., Fels, D.I.: Gamification in theory and action: A survey.
  International Journal of Human-Computer Studies  \textbf{74},  14--31 (Feb
  2015). \doi{10.1016/j.ijhcs.2014.09.006},
  \url{https://doi.org/10.1016/j.ijhcs.2014.09.006}

\bibitem{10.1145/2934687}
Srba, I., Bielikova, M.: A comprehensive survey and classification of
  approaches for community question answering. ACM Trans. Web  \textbf{10}(3)
  (aug 2016). \doi{10.1145/2934687}, \url{https://doi.org/10.1145/2934687}

\bibitem{thayer2004localization}
Thayer, A., Kolko, B.E.: Localization of digital games: The process of blending
  for the global games market. Technical communication  \textbf{51}(4),
  477--488 (2004)

\bibitem{10.1007/978-3-030-29384-0_23}
Tondello, G.F., Arrambide, K., Ribeiro, G., Cen, A.J.l., Nacke, L.E.: ``i don't
  fit into a single type'': A trait model and scale of game playing
  preferences. In: Lamas, D., Loizides, F., Nacke, L., Petrie, H., Winckler,
  M., Zaphiris, P. (eds.) Human-Computer Interaction -- INTERACT 2019. pp.
  375--395. Springer International Publishing, Cham (2019)

\bibitem{10.1145/2967934.2968082}
Tondello, G.F., Wehbe, R.R., Diamond, L., Busch, M., Marczewski, A., Nacke,
  L.E.: The gamification user types hexad scale. In: Proceedings of the 2016
  Annual Symposium on Computer-Human Interaction in Play. p. 229–243. CHI
  PLAY '16, Association for Computing Machinery, New York, NY, USA (2016).
  \doi{10.1145/2967934.2968082}, \url{https://doi.org/10.1145/2967934.2968082}

\bibitem{10.1145/3313831.3376755}
Troiano, G.M., Chen, Q., Alba, A.V., Robles, G., Smith, G., Cassidy, M.,
  Tucker-Raymond, E., Puttick, G., Harteveld, C.: Exploring how game genre in
  student-designed games influences computational thinking development. In:
  Proceedings of the 2020 CHI Conference on Human Factors in Computing Systems.
  p. 1–17. CHI '20, Association for Computing Machinery, New York, NY, USA
  (2020). \doi{10.1145/3313831.3376755},
  \url{https://doi.org/10.1145/3313831.3376755}

\bibitem{10.1145/3313831.3376723}
Tyack, A., Mekler, E.D.: Self-determination theory in hci games research:
  Current uses and open questions. In: Proceedings of the 2020 CHI Conference
  on Human Factors in Computing Systems. p. 1–22. CHI '20, Association for
  Computing Machinery, New York, NY, USA (2020). \doi{10.1145/3313831.3376723},
  \url{https://doi.org/10.1145/3313831.3376723}

\bibitem{10.1145/3311350.3347182}
Urbanek, M., G\"{u}ldenpfennig, F.: Unpacking the audio game experience:
  Lessons learned from game veterans. In: Proceedings of the Annual Symposium
  on Computer-Human Interaction in Play. p. 253–264. CHI PLAY '19,
  Association for Computing Machinery, New York, NY, USA (2019).
  \doi{10.1145/3311350.3347182}, \url{https://doi.org/10.1145/3311350.3347182}

\end{thebibliography}
%




\end{document}